\documentclass[aip,physrev,reprint,groupedaddress]{revtex4-2}
\usepackage[most]{tcolorbox}
\usepackage{physics}

\usepackage{tikz}
\usetikzlibrary{decorations.pathmorphing}

\begin{document}

\title{Noise-induced decoherence-free zones for anyons}

\author{Eric~R.~Bittner}
\email{ebittner@central.uh.edu}
%https://orcid.org/0000-0002-0775-9664
\affiliation{Department of Physics, University of Houston, Houston, Texas 77204, United~States}

\date{\today}

\begin{abstract}
We develop a stochastic framework for anyonic systems in which the exchange
phase is promoted from a fixed parameter to a fluctuating quantity. Starting
from the Stratonovich stochastic Liouville equation, we perform the
Stratonovich--Itô conversion to obtain a Lindblad master equation that ties the
dissipator directly to the distorted anyon algebra. This construction produces
a statistics--dependent dephasing channel, with rates determined by the
eigenstructure of the real symmetric correlation matrix $D_{ab}$. The
eigenvectors of $D$ select which collective exchange currents---equivalently,
which irreducible representations of the system---are protected from stochastic
dephasing, providing a natural mechanism for decoherence-free subspaces and
noise-induced exceptional points. The key result of our analysis is the
universality of the optimal statistical angle: in the minimal two-site model
with balanced gain and loss, the protected mode always minimizes its dephasing
at $\theta^\star = \pi/2$, independent of the specific form of $D$. This
robustness highlights a simple design rule for optimizing coherence in noisy
anyonic systems, with direct implications for ultracold atomic realizations and
other emerging platforms for fractional statistics.
\end{abstract}

%\keywords{}

\maketitle

\section{Introduction}

Quantum statistics provides the fundamental classification of indistinguishable
particles. In three spatial dimensions, particle exchange produces only two
consistent outcomes: bosons, which acquire a trivial phase $e^{i0}=+1$, and
fermions, which acquire a minus sign $e^{i\pi}=-1$. This restriction follows
from the topology of configuration space in 3D, where exchange paths can be
continuously deformed into one another and only the two representations of the
permutation group are allowed.

In reduced dimensionality, the situation changes dramatically. In two
dimensions the braid group replaces the permutation group, and continuous
families of representations become possible. Leinaas and Myrheim first showed
that indistinguishable particles in two dimensions can obey generalized
exchange statistics interpolating between bosons and fermions
\cite{Leinaas1977}, while Wilczek introduced the term \emph{anyon} and provided
explicit models for such particles \cite{Wilczek1982}. Fractional quantum Hall
states offered the first concrete physical setting in which such excitations
arise, with Laughlin’s wavefunction \cite{Laughlin1983} and the Arovas--Schrieffer--Wilczek
analysis of Berry phases \cite{Arovas1984} establishing the connection between
fractional charge and fractional statistics. Extensions to one-dimensional
systems further revealed that constrained motion can effectively realize
anyon-like exchange \cite{Haldane1991,Kundu1999}. Anyons are now recognized as
central to our understanding of topological phases of matter, from spin liquids
\cite{Kitaev2003} to non-Abelian braiding proposals for quantum computation
\cite{Nayak2008}.

Experimental realizations have advanced dramatically. For many years,
experimental evidence for anyons was restricted to indirect signatures in
two-dimensional electron gases. Interferometric probes in the fractional
quantum Hall regime have provided strong support for fractional statistics,
while more recent work has advanced toward direct braiding experiments. A
breakthrough came with the realization of one-dimensional anyons in ultracold
atomic systems: Kwan \emph{et al.} engineered a density-dependent Peierls phase
in an optical lattice and demonstrated control of an arbitrary statistical
angle, confirming anyonic behavior through quantum walks, Hanbury
Brown--Twiss interference, and bound-state formation in the two-particle sector
\cite{Kwan2024Science}. These advances establish that not only can the mean
statistical phase be tuned, but anyons can be probed in platforms with high
degrees of controllability.

In realistic settings, the statistical phase is never perfectly fixed.
Environmental coupling, microscopic disorder, or engineered modulation can
introduce fluctuations about the mean value of $\theta$. These fluctuations
render the effective exchange factor stochastic,
\begin{equation}
  e^{i\theta} \;\longrightarrow\; e^{i(\theta+\phi(t))},
\end{equation}
with $\phi(t)$ encoding dynamical noise. Our central goal in this work is to
develop a consistent framework for describing anyons subject to such
fluctuating exchange phases.

Recent theoretical work has also shown that noise can play a constructive role
in quantum dynamics. In particular, we have demonstrated that correlated noise
can drive phase synchronization between otherwise independent quantum systems
\cite{Bittner2024,Bittner2024b}, and related studies have emphasized the broader
importance of correlation structure in open-system dynamics \cite{Hou2025prr}.
These works highlight how the form of the noise correlation matrix can select
protected collective modes and suppress decoherence. The present work extends this line of research into the realm of anyonic statistics, showing that
fluctuations of the statistical phase lead to analogous protection mechanisms
and, under suitable conditions, to noise-induced exceptional points. In this
way, the framework developed here builds directly on our synchronization
results while situating them within the emerging literature on correlated noise
in quantum systems.

In this paper we formulate this problem by assigning each tunneling link
between sites a stochastic phase variable $\phi_a(t)$. Starting from the
Stratonovich stochastic Liouville equation, we perform the Stratonovich--Itô
conversion to obtain a Lindblad master equation. This yields a
statistics--dependent pure dephasing channel of the form
$-\tfrac{\Gamma_\theta}{2}[K_\theta,[K_\theta,\rho]]$, where the
exchange--current operator $K_\theta$ encodes the distorted anyon algebra. From
this starting point, we obtain several new results. First, we demonstrate that
stochastic exchange phases generate a dephasing channel whose rate depends
explicitly on the statistical angle. Second, in the minimal two-site
broken-$\mathcal{PT}$ model, we show that the protected eigenmode is
universally stabilized at an optimal angle $\theta^\star=\pi/2$, where the
dephasing channel vanishes in the absence of residual relaxation. Third, for
multiple noisy links, correlations between phase fluctuations are captured by a
correlation matrix $D_{ab}$. Real-symmetric correlations yield collective
exchange currents and decoherence-free subspaces when $D$ loses rank. Finally,
we establish that noise-induced exceptional points cannot arise for
real-symmetric $D$, but become possible when $D$ carries complex or chiral
correlations, which render the dissipator non-normal.

Taken together, these results provide a systematic framework for understanding
how fluctuating statistical phases affect anyonic coherence and protection. By
tying stochastic dephasing directly to the underlying anyon algebra, we offer a
set of design rules for engineering ``designer phases'' in noisy anyonic
systems, with relevance for ultracold atomic realizations and other platforms
where fractional statistics are emerging as experimentally accessible degrees
of freedom.

\section{Theory}

\subsection{From Distorted Anyon Algebra to the Model Hamiltonian}

To motivate our model Hamiltonian we begin with the distorted algebra that defines
abelian anyons. In a fixed site ordering ($1<2<\cdots$), the annihilation operators
obey
\begin{equation}
a_i a_j = e^{i\theta}\,a_j a_i, \qquad 
a_i a_j^\dagger = e^{-i\theta}\,a_j^\dagger a_i \quad (i<j),
\label{eq:anyonalgebra}
\end{equation}
with $\theta$ the anyonic statistical angle. This algebra can be realized by
bosonic operators dressed with Jordan--Wigner strings,
\begin{equation}
a_j = b_j \exp\!\Big(i\theta \sum_{k<j} n_k\Big), \qquad n_k=b_k^\dagger b_k,
\end{equation}
so that exchanging two particles produces the phase factor $e^{i\theta}$.

For two sites, labeled 1 and 2, the exchange operator that transfers an excitation
from site 2 to site 1 carries this intrinsic anyonic phase,
\begin{equation}
\mathcal T_\theta = a_1^\dagger a_2\, e^{i\theta},
\end{equation}
and the associated Hermitian exchange current is
\begin{equation}
K_\theta = i\!\left(\mathcal T_\theta - \mathcal T_\theta^\dagger\right)
= i\!\left(a_1^\dagger a_2 e^{i\theta} - a_2^\dagger a_1 e^{-i\theta}\right).
\label{eq:Ktheta}
\end{equation}

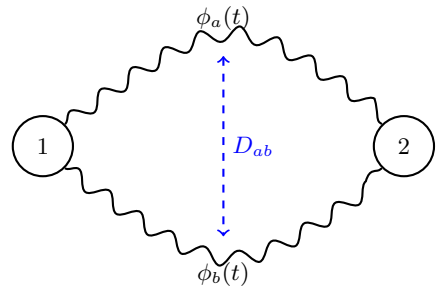
\begin{figure}[t]
\centering
\begin{tikzpicture}[scale=1.2, thick]

  % Nodes
  \node[circle,draw,minimum size=0.8cm] (A) at (0,0) {1};
  \node[circle,draw,minimum size=0.8cm] (B) at (4,0) {2};

  % Upper noisy path
  \draw[decorate,decoration={snake,amplitude=1mm,segment length=5mm}] 
    (A.north east) .. controls (2,1.5) .. (B.north west)
    node[midway,above] {$\phi_a(t)$};

  % Lower noisy path
  \draw[decorate,decoration={snake,amplitude=1mm,segment length=5mm}] 
    (A.south east) .. controls (2,-1.5) .. (B.south west)
    node[midway,below] {$\phi_b(t)$};

  % Correlation (dashed)
  \draw[dashed,<->,blue] (2,1.0) -- (2,-1.0)
    node[midway,right,blue] {$D_{ab}$};

\end{tikzpicture}
\caption{Two sites connected by fluctuating paths with random phases 
$\phi_a(t)$ and $\phi_b(t)$. A correlation $D_{ab}$ between 
the two noise sources captures the strength of correlated environmental 
fluctuations.}
\label{fig:1}
\end{figure}

In a realistic environment, the statistical phase imprinted on each tunneling link
is not fixed. Each physical path may accumulate an additional stochastic phase
$\phi_a(t)$ due to environmental fluctuations or engineered modulation. This amounts
to promoting the statistical factor $e^{i\theta}$ to a stochastic link variable
$e^{i(\theta+\phi_a(t))}$,
\begin{equation}
\mathcal T_\theta \;\longrightarrow\; \mathcal T_\theta\,e^{i\phi_a(t)}
= a_1^\dagger a_2\,e^{i(\theta+\phi_a(t))}.
\end{equation}
For multiple paths $a,b,\dots$, each link carries its own $\phi_a(t)$, with correlations
encoded by the diffusion (or quantum noise) matrix
\(
\langle \dd\phi_a \dd\phi_b \rangle = 2D_{ab}\,\dd t.
\)

We therefore arrive at the physical picture illustrated schematically in Fig.~\ref{fig:1} which 
illustrates 
two sites coupled by two noisy tunneling channels, each carrying both the intrinsic
anyonic statistical phase $\theta$ and extrinsic stochastic components $\phi_a(t)$ and $\phi_b(t)$ 
which may be correlated via $D_{ab}$.
If multiple tunneling paths $a,b,\cdots$ connect sites $1$ and $2$, each
contributes its own exchange operator $\mathcal T_\theta^{(a)}$ and stochastic
phase $\phi_a(t)$. 
The statistics of the stochastic phases are characterized by the correlation
matrix
\begin{equation}
\langle \dd\phi_a \dd\phi_b\rangle = 2 D_{ab}\,\dd t ,
\end{equation}
which encodes how noise on different links is correlated. The structure of
$D_{ab}$ will determine the collective dephasing modes of the system, as
we analyze in the following sections.

The effective Hamiltonian is then
\begin{equation}
H(t) = H_0 - \sum_a J_a\!\left(\mathcal T_\theta^{(a)} e^{i\phi_a(t)} 
+ \mathcal T_\theta^{(a)\dagger} e^{-i\phi_a(t)}\right),
\label{eq:multi-link-H}
\end{equation}
where $J_a$ is the tunneling amplitude along link $a$. Each link therefore
carries both the intrinsic anyonic statistical phase $e^{i\theta}$ and an
extrinsic stochastic modulation $e^{i\phi_a(t)}$.

For the case of a single exchange path, 
we have the effective Hamiltonian
\begin{equation}
H(t) = H_0 \;-\; J\!\left(\mathcal T_\theta\, e^{i\phi(t)} + \mathcal T_\theta^\dagger e^{-i\phi(t)}\right),
\qquad 
\mathcal T_\theta = a_1^\dagger a_2\,e^{i\theta},
\end{equation}
where $H_0$ contains the local mode energies and $\theta$ is the fixed anyonic statistical phase.

For small and rapidly fluctuating $\phi(t)$ we expand the exponential to linear order and collect terms into the Hermitian exchange--current operator.
The Hamiltonian then takes the approximate form
\begin{equation}
H(t) \simeq H_0 - J K_\theta \,\phi(t).
\end{equation}

We assume $\phi(t)$ undergoes phase diffusion according to a stochastic differential equation (SDE)
such as 
\begin{equation}
\dd \phi(t) = \sqrt{2D_\phi}\,\dd W_t, 
\end{equation}
where $W_t$ is a standard Wiener process and $D_\phi$ is the phase--diffusion constant.  
The associated Stratonovich stochastic Liouville equation becomes
\begin{equation}
\dd \rho = -\,i[H_0,\rho]\,\dd t \;-\; iJ\,[K_\theta,\rho]\circ \dd \phi(t)
\label{eq:stoch_strat}
\end{equation}
where the symbol $\circ$ denotes Stratonovich integration.
We distinguish Itô from Stratonovich stochastic calculus by notation: 
$\dd X = A\,\dd t + B\,\dd W_t$ denotes the Itô form, while 
$\dd X = A\,\dd t + B\circ\dd W_t$ indicates the Stratonovich form, 
with the symbol $\circ$ specifying the Stratonovich interpretation.

To connect with ensemble-averaged dynamics we convert Eq.~\eqref{eq:stoch_strat} into Itô form.  
For a Stratonovich stochastic differential equation
\begin{equation}
\dd \rho = \mathcal A(\rho)\,\dd t + \mathcal G(\rho)\circ \dd W_t,
\end{equation} the corresponding Itô form reads
\begin{equation}
\dd \rho = \Big[\mathcal A(\rho) + \tfrac12 \,\mathcal G'(\rho)\!\cdot\!\mathcal G(\rho)\Big]\dd t 
+ \mathcal G(\rho)\,\dd W_t,
\end{equation}
where $\mathcal G'$ is the Fréchet derivative.  
In our case, the noise superoperator is linear in $\rho$,
\begin{equation}
\mathcal G(\rho) = \sqrt{2D_\phi}\,(-iJ)\,[K_\theta,\rho].
\end{equation}
Hence $\mathcal G'(\rho)\!\cdot\!\mathcal X = \mathcal G(\mathcal X)$ and
\begin{equation}
\tfrac12 \,\mathcal G(\mathcal G(\rho)) 
= -J^2 D_\phi\,[K_\theta,[K_\theta,\rho]].
\end{equation}

The full Itô equation of motion is therefore
\begin{align}
\dd \rho &= \Big(-\,i[H_0,\rho] - J^2 D_\phi [K_\theta,[K_\theta,\rho]]\Big)\dd t\nonumber \\
&+ \sqrt{2D_\phi}\,(-iJ)[K_\theta,\rho]\,\dd W_t.
\label{eq:ito_full}
\end{align}

Averaging over the stochastic increments removes the explicit noise term in Eq.~\eqref{eq:ito_full}, leaving
\begin{equation}
\dot \rho = -\,i[H_0,\rho] \;-\; \frac{\Gamma_\theta}{2}\,[K_\theta,[K_\theta,\rho]],
\qquad \Gamma_\theta = 2J^2 D_\phi.
\label{eq:master}
\end{equation}
This is a deterministic master equation in Lindblad form with a single Hermitian jump operator,
\begin{equation}
L_\theta = \sqrt{\Gamma_\theta}\,K_\theta.
\end{equation}
Random exchange phases therefore produce pure dephasing in the eigenbasis of $K_\theta$, with rates that depend explicitly on the anyonic statistical angle $\theta$.  

As an aside, the approach generalizes to other noise models.
For instance, if the random phase follows an Ornstein--Uhlenbeck process with correlation function $C_\phi(\tau)=\sigma^2 e^{-|\tau|/\tau_c}$, the white-noise limit yields
\begin{equation}
\Gamma_\theta = 2J^2\int_0^\infty C_\phi(\tau)\,\dd\tau = 2J^2\sigma^2\tau_c
\end{equation}
where $\sigma$ governs the gaussian width of 
fluctuations (classically proportional to temperature) and $\tau$ is the bath correlation time. 
We also note that the random phase can also be treated as 
an operator-valued phase generated by a quantum bath.
In this case we still obtain the Lindblad form, however, 
with a rate proportional to the zero-frequency 
part of the noise-spectrum.

In practice, however, quasiparticles also undergo 
population relaxation due to coupling to thermal baths or lossy reservoirs.  
The combined dynamics can be written schematically as
\begin{equation}
\dot\rho = -\,i[H_0,\rho]
\;-\;\frac{\Gamma_\theta}{2}[K_\theta,[K_\theta,\rho]]
\;+\;\sum_\alpha \gamma_\alpha\,\mathcal D[L_\alpha]\rho,
\end{equation}
where the last term collects Lindblad dissipators
$\mathcal{D}[L]\rho = L\rho L^\dagger - \tfrac12\{L^\dagger L,\rho\}$,
which generate population relaxation (and, for thermal baths, detailed-balance
thermalization). The essential physics is the \emph{competition} between the
statistics–dependent pure dephasing channel and the relaxation channel. The
former damps coherences without altering populations, while the latter reshuffles
populations and sets lifetime broadening. This same competition lies at the
heart of the synchronization mechanism analyzed in our recent work on noisy
anyonic systems: correlations in the stochastic phases can suppress one channel
relative to the other, thereby stabilizing collective phase locking. In the
$\mathcal{PT}$-symmetric regime this balance yields robust coherent oscillations,
whereas in the broken phase the non-orthogonality of modes amplifies both
mechanisms and accelerates decoherence.
Pure dephasing suppresses coherences without altering populations, 
while relaxation reshuffles populations and can either restore or 
destabilize coherence depending upon the symmetry of the jump operators.  

\subsection{Decoherence free states}

Having established that stochastic exchange phases give rise to a
statistics--dependent dephasing channel in direct competition with
population relaxation, we now turn to the conditions under which
coherence can persist. Of particular interest are situations where
the system supports modes that are insensitive to the noisy exchange
phases. These decoherence-free states are selected by the structure
of the exchange-current operator $K_\theta$ and by correlations among
different tunneling paths. Identifying such protected modes allows
us to quantify both their lifetimes and the statistical angles at
which they are optimally stabilized.

To explore this, let us suppose there exists a 
single mode that is protected from the noise, $\ket{u}$,
and that we initiate the system in that state as a 
pure state with density operator 
$\rho_u(0)=\ket{u}\!\bra{u}$.  Even when relaxation is suppressed, the stochastic exchange-phase 
channel derived above remains active and imposes a finite lifetime of this protected mode. For this, we evaluate the instantaneous decay of the survival probability 
$s_u(t)=\tr(\rho_u(t)\rho_u(0))$ at $t=0$:
\begin{align}
\left.\frac{\dd}{\dd t}\, s_u(t)\right|_{t=0}
&= \tr\!\left(\rho_u\,\dot\rho\right)
= -\,\Gamma_\theta\,\mathrm{Var}_{\ket{u}}\!\left(K_\theta\right),\nonumber \\
\mathrm{Var}_{\ket{u}}(K_\theta)
&:= \bra{u}K_\theta^2\ket{u}-\bra{u}K_\theta\ket{u}^2 .
\label{eq:variance-rule}
\end{align}
Thus, the \emph{effective decoherence rate} of the  mode is
\begin{equation}
\gamma_{\phi,u}(\theta) \;=\; \Gamma_\theta\,\mathrm{Var}_{\ket{u}}\!\left(K_\theta\right),
\qquad 
\tau(\theta) \;=\; \gamma_{\phi}(\theta)^{-1} ,
\label{eq:tau-variance}
\end{equation}
which depends on the statistical angle exclusively through the exchange-current operator 
$K_\theta$.

\begin{figure}[t]
  \centering
  \includegraphics[width=\columnwidth]{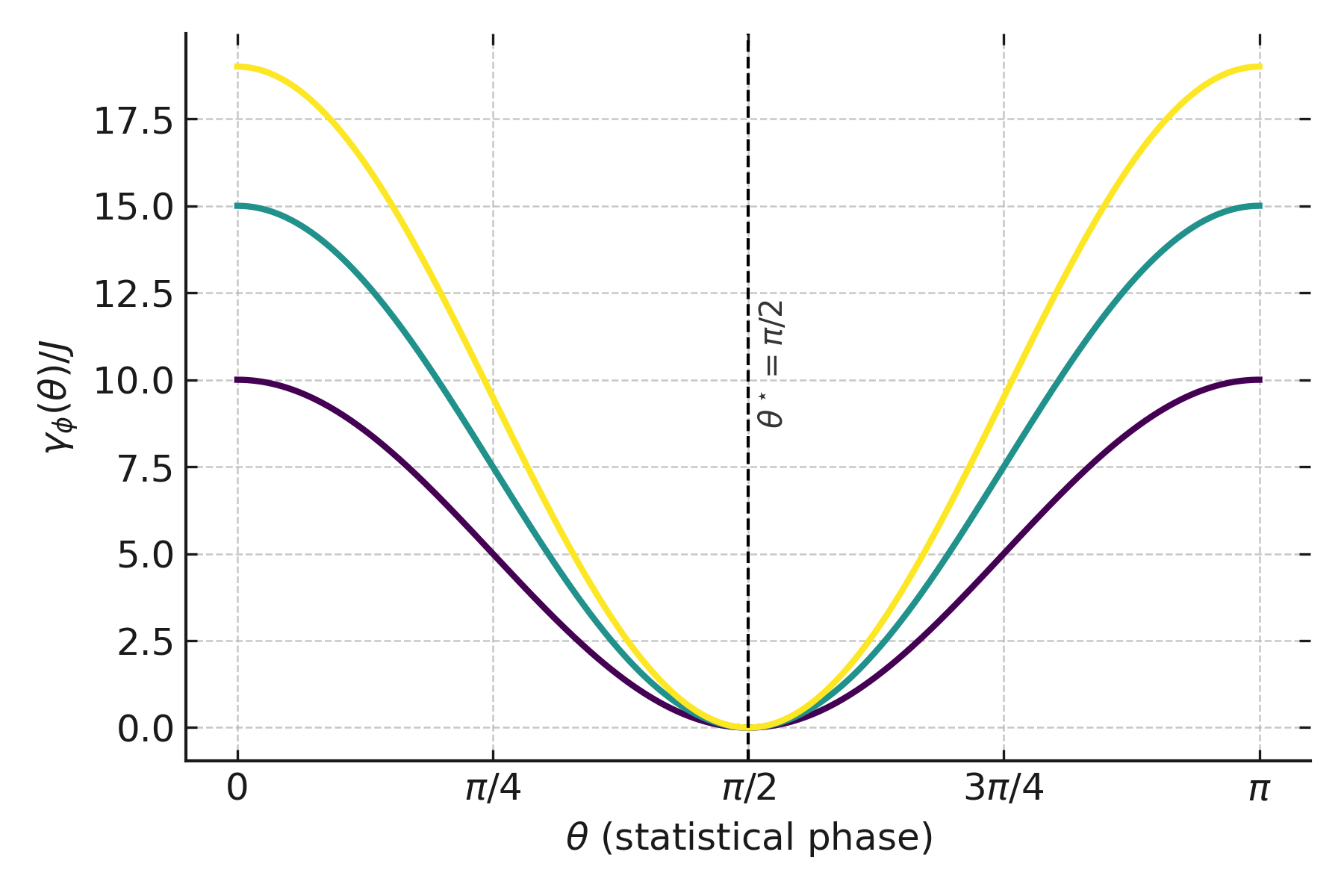}
  \caption{
  Dephasing rate scaled by exchange coupling, $\gamma_\phi(\theta)/J$, versus statistical phase $\theta$
  for three representative correlation strengths:
  \textcolor[rgb]{0.267004,0.004874,0.329415}{\rule[2pt]{10pt}{1pt}} $\xi=0$,
  \textcolor[rgb]{0.127568,0.566949,0.550556}{\rule[2pt]{10pt}{1pt}} $\xi=0.5$, and
  \textcolor[rgb]{0.993248,0.906157,0.143936}{\rule[2pt]{10pt}{1pt}} $\xi=0.9$.
  All curves exhibit a universal minimum at $\theta^\star = \pi/2$, independent of $\xi$,
  demonstrating the robustness of the half-fermionic protection point. Rates are expressed
  in dimensionless units normalized by the exchange coupling $J$ (with $J=0.1$ in the
  underlying simulation).
  }
  \label{fig:gamma-over-J}
\end{figure}

The statistics-dependent dephasing rate $\gamma_\phi(\theta)$ exhibits a simple and universal angular dependence.
In Fig.~\ref{fig:gamma-over-J}, we plot the normalized rate $\gamma_\phi(\theta)/J$ as a function of the statistical phase $\theta$ for several values of the noise correlation coefficient $\xi$.
Although the overall magnitude and curvature of $\gamma_\phi(\theta)$ depend on the degree of interlink correlation, the location of the minimum remains fixed at $\theta^\star = \pi/2$.
This point corresponds to the half-fermionic statistics, where the exchange-current operator $K_\theta$ becomes orthogonal to the protected mode, minimizing the variance $\mathrm{Var}_{|u\rangle}(K\theta)$ and hence the dephasing.
The invariance of $\theta^\star$ with respect to $\xi$ demonstrates that the protection mechanism is purely algebraic and not sensitive to the detailed structure of the noise correlation matrix $D_{ab}$.
All rates are expressed in dimensionless units scaled by the exchange coupling $J$, with $J=0.1$ in the simulations shown.

\subsection{Two-mode single-excitation manifold.}
In the one-particle subspace spanned by $\{\ket{1},\ket{2}\}$, the exchange-current operator takes the form
\begin{equation}
K_\theta \;=\; -\sin\theta\,\sigma_x \;+\; \cos\theta\,\sigma_y ,
\qquad 
K_\theta^2=\openone,
\end{equation}
so that $\mathrm{Var}_{\ket{u}}(K_\theta) = 1 - \big(\mathbf{n}(\theta)\!\cdot\!\mathbf{r}_u\big)^2$,
where $\mathbf{n}(\theta)=(-\sin\theta,\cos\theta,0)$ and 
$\mathbf{r}_u=(\bra{u}\sigma_x\ket{u},\bra{u}\sigma_y\ket{u},\bra{u}\sigma_z\ket{u})$ 
is the Bloch vector of the initial state.  Substituting into Eq.~\eqref{eq:tau-variance} yields
\begin{equation}
\gamma_{\phi,u}(\theta) \;=\; \Gamma_\theta \left[\,1 - \big(\mathbf{n}(\theta)\!\cdot\!\mathbf{r}_u\big)^2\right].
\label{eq:gamma-bloch}
\end{equation}
We therefore \emph{maximize} the lifetime by maximizing $|\mathbf{n}(\theta)\!\cdot\!\mathbf{r}_u|$, i.e.\ by aligning 
$\mathbf{n}(\theta)$ with the in-plane projection of $\mathbf{r}_u$.  
Writing $\mathbf{r}_u^{\parallel}=(r_x,r_y,0)$ and $\arg(r_x+ i r_y)=:\varphi_u$, the optimal statistical angle is
\begin{equation}
{\;\theta^\star \;=\; \frac{\pi}{2}-\varphi_u \;\;(\mathrm{mod}\;\pi)\;, \qquad
\gamma_{\phi,u}^{\min} \;=\; \Gamma_\theta\!\left[1 - \|\mathbf{r}_u^{\parallel}\|^2\right].\;}
\label{eq:theta-star}
\end{equation}
Physically, we choose $\theta$ so that $\ket{u}$ is \emph{as nearly an eigenstate of $K_\theta$ as possible}.  
If $\ket{u}$ is exactly an eigenstate of $K_\theta$ (so $\|\mathbf{r}_u^{\parallel}\|=1$ and 
$\mathbf{n}(\theta^\star)\!\parallel\!\mathbf{r}_u^{\parallel}$), then 
$\mathrm{Var}_{\ket{u}}(K_{\theta^\star})=0$ and the stochastic exchange-phase channel does not decohere the protected mode. Thus, the lifetime of the 
mode becomes 
\emph{infinite} in the ideal absence of relaxation.

This is a significant conclusion: \emph{in the simplest two--mode broken-$\mathcal{PT}$ model, 
the optimal statistical angle collapses to a universal value of $\theta^\star=\pi/2$}.  
At this angle the exchange--current operator reduces to
\begin{equation}
K_{\pi/2} = -\sigma_x,
\end{equation}
so that the protected mode is effectively an eigenstate of $K_\theta$ and 
the stochastic exchange--phase channel cannot induce decoherence.  
The corresponding lifetime $\tau_{\mathrm{eff},u}(\theta^\star)$ 
diverges in the absence of residual relaxation.

This result is robust even in the presence of 
residual relaxation channels as we can see by 
simply adding an additional $\gamma_{res}$ to the 
relaxation rate, 
\begin{align}
\gamma_{\mathrm{eff},u}(\theta) \;&=\; \gamma_{\mathrm{res}} \;+\; \Gamma_\theta\,\mathrm{Var}_{\ket{u}}\!\left(K_\theta\right).
\end{align}
Hence, the lifetime is given by 
\begin{align}
\tau_{\mathrm{eff},u}(\theta) \;&=\; \big[\gamma_{\mathrm{res}} + \Gamma_\theta(1-(\mathbf{n}\!\cdot\!\mathbf{r}_u)^2)\big]^{-1}.
\end{align}
Since our optimization maximizes $\Gamma_\theta\,\mathrm{Var}_{\ket{u}}\!\left(K_\theta\right)$, the optimal statistical angle $\theta^\star$ is still given by Eq.~\eqref{eq:theta-star}.

The universality of $\theta^\star=\pi/2$ stems from the real structure of the 
protected eigenmode in the broken-$\mathcal{PT}$ regime: the absence of an 
intrinsic phase between site amplitudes forces alignment with the $\sigma_x$ 
axis.  Nontrivial dependence of $\theta^\star$ on system parameters requires 
a complex Bloch vector with $r_y\neq 0$, which can arise from asymmetric 
couplings or coupling to a chiral (complex--correlated) bath.  
Thus, while the universal angle $\pi/2$ reflects the robustness of the minimal model, 
it also highlights a route for engineering tunable protection through 
environmental chirality and correlation.

% In the $\mathcal{PT}$--symmetric regime, these two processes may balance 
% to yield stable oscillations with finite coherence lifetimes.  
% In contrast, once $\mathcal{PT}$ symmetry is broken, the non--orthogonality 
% of the system eigenmodes amplifies noise and leads to rapid decay of both 
% coherence and population.  
% Exceptional points thus mark the boundaries where dephasing and relaxation 
% interplay most strongly, producing qualitatively new dynamical behavior.

% From the spectroscopic perspective, this competition directly controls 
% line broadening and the survival of higher--order coherences.  
% Pure dephasing contributes homogeneous linewidths that scale with $\Gamma_\theta$, 
% while relaxation adds lifetime broadening that scales with the $\gamma_\alpha$.  
% In broken $\mathcal{PT}$ phases, these broadenings do not add linearly; 
% rather, they interfere due to mode non--orthogonality, yielding 
% nontrivial linewidth scaling and strongly asymmetric line shapes.  
% This interplay therefore provides a direct experimental fingerprint 
% of the stochastic--phase dynamics in the vicinity of exceptional points.

\paragraph{Correlated links / extended graphs.}
For systems with multiple links $a$ carrying exchange-current operators $K_\theta^{(a)}$ and a positive semidefinite  phase-noise matrix $\Gamma_{ab}$, the protected-mode dephasing generalizes to
\begin{equation}
\gamma_{\phi,u}(\theta) \;=\; 
\frac{1}{2}\sum_{a,b}\Gamma_{ab}\,\mathrm{Cov}_{\ket{u}}\!\big(K_\theta^{(a)},K_\theta^{(b)}\big),
\end{equation}
in which the covariance between channels is given by
\begin{equation}
\mathrm{Cov}_{\ket{u}}(A,B)
:= \bra{u}\tfrac{1}{2}\{A,B\}\ket{u} - \bra{u}A\ket{u}\bra{u}B\ket{u}.
\label{eq:cov-form}
\end{equation}
Equation~\eqref{eq:cov-form} suggests that correlated phase noise can \emph{reduce} the protected-mode dephasing when 
the covariances interfere destructively, providing an additional design knob complementary to the $\theta$ 
statistical angle. 

% In the broken-$\mathcal{PT}$ phase, the protected mode’s lifetime is limited by 
% stochastic exchange-phase dephasing at rate 
% $\gamma_{\phi,u}(\theta)=\Gamma_\theta\,\mathrm{Var}_{\ket{u}}(K_\theta)$.
% \emph{Choose the statistical angle $\theta$ that aligns $K_\theta$ with the protected mode}, 
% so that $\ket{u}$ is (as nearly as possible) an eigenstate of $K_\theta$.  
% In the two-mode single-excitation manifold, this yields 
% $\theta^\star=\tfrac{\pi}{2}-\arg(r_x+i r_y)$, with a minimal rate 
% $\gamma_{\phi,u}^{\min}=\Gamma_\theta\!\left[1-\|\mathbf{r}_u^{\parallel}\|^2\right]$.  
% Correlated phase noise across links can further \emph{cancel} dephasing via covariance interference 
% as in Eq.~\eqref{eq:cov-form}.

\subsection{Correlated Phase Noise on Two Links: Rates, DFS, and (the Absence of) Decoherence EPs}

We now consider two links $a=1,2$ with exchange--current operators $K_\theta^{(1)}$ and $K_\theta^{(2)}$.
Correlated stochastic exchange phases $\{\phi_a(t)\}$ obey
\begin{equation}
\langle \dd \phi_a \dd \phi_b\rangle = 2\,\mathsf D_{ab}\,\dd t,
\qquad 
\mathsf D \;=\; \begin{pmatrix}1 & \xi \\[2pt] \xi & 1\end{pmatrix},\quad |\xi|\le 1,
\end{equation}
and we take equal tunneling amplitudes for clarity ($J_1=J_2=J$).  
Ignoring population relaxation, the ensemble--averaged Liouvillian reads
\begin{equation}
\dot\rho \;=\; -\,i[H_0,\rho]
- \frac{1}{2}\sum_{a,b=1}^2 \Gamma_{ab}\,[K_\theta^{(a)},[K_\theta^{(b)},\rho]],
\label{eq:two-link-L}
\end{equation}
with
\begin{equation}
\Gamma_{ab} = 2J^2 \mathsf D_{ab}.
\end{equation}

Define symmetric/antisymmetric combinations $K_\theta^{(\pm)}=(K_\theta^{(1)}\pm K_\theta^{(2)})/\sqrt{2}$.  
Since $\mathsf D$ is real symmetric and positive semidefinite, it diagonalizes as
\begin{equation}
\mathsf D = U\,\mathrm{diag}(1+\xi,\,1-\xi)\,U^\top,
\qquad 
U=\frac{1}{\sqrt2}\begin{pmatrix}1&1\\[2pt]1&-1\end{pmatrix}.
\end{equation}
Equation~\eqref{eq:two-link-L} becomes a \emph{sum of two independent dephasers}:
\begin{equation}
\dot\rho = -\,i[H_0,\rho]
-\frac{\gamma_+}{2}\,[K_\theta^{(+)},[K_\theta^{(+)},\rho]]
-\frac{\gamma_-}{2}\,[K_\theta^{(-)},[K_\theta^{(-)},\rho]],
\end{equation}
with 
\begin{equation}
\gamma_\pm \;=\; 2J^2(1\pm\xi).
\label{eq:rates-pm}
\end{equation}
Thus correlated phases simply select two \emph{collective} exchange currents with rates $\gamma_\pm$.

At $\xi=+1$ we have $\gamma_-=0$ while $\gamma_+=4J^2$: the antisymmetric channel $K_\theta^{(-)}$ is noiseless, and any operator commuting with $K_\theta^{(+)}$ forms a decoherence--free subspace (DFS).  
At $\xi=-1$ the roles swap ($\gamma_+=0$, $\gamma_-=4J^2$). 
These are \emph{rank--deficiency} points of $\mathsf D$ (and $\Gamma$), where one stochastic eigenmode is completely suppressed.  
They are \emph{not} exceptional points (EPs) in the spectral sense; the Liouvillian remains diagonalizable.
This occurs because the generator \eqref{eq:two-link-L} is a sum of \emph{double commutators with Hermitian operators}.  
Equivalently, it is a sum of Lindblad dissipators with \emph{Hermitian} jump operators:
\begin{equation}
\mathcal L[\rho] = -\,i[H_0,\rho] 
+ \sum_{\nu=\pm} \gamma_\nu\!\left( K_\theta^{(\nu)} \rho K_\theta^{(\nu)} - \tfrac12\{(K_\theta^{(\nu)})^2,\rho\} \right).
\end{equation}
Such a Liouvillian is \emph{normal} with respect to the Hilbert--Schmidt inner product; its spectrum is real and it is diagonalizable by construction.  
Degeneracies of decay rates (e.g.\ $\gamma_+=\gamma_-$ at $\xi=0$) do not create nontrivial Jordan blocks.  
Hence, with real symmetric $\mathsf D$ and Hermitian $K$’s, one does \emph{not} obtain a spectral EP; instead, one encounters DFS formation when $\mathrm{rank}(\mathsf D)$ drops.

In contrast, when the environmental noise correlates the relaxation
channels of distinct sites, the effective Lindblad operators couple
collective annihilation modes rather than acting locally. A minimal
realization is
\[
L_\pm = \tfrac{1}{2}\sqrt{1\pm\xi}\,\sqrt{\gamma}\,(a_1 \pm a_2),
\]
where $\xi$ quantifies the degree of correlation between the local
dissipative baths. These non-local jump operators mix the two sites and
render the full Liouvillian \emph{non-normal} under the
Hilbert--Schmidt inner product, even though each $L_\pm$ remains
annihilation-like. As $\xi$ is tuned, the Liouvillian eigenvalues can
coalesce together with their corresponding eigenmodes, giving rise to
\emph{decoherence exceptional points} (EPs) that separate synchronized
and desynchronized dynamical regimes. This mechanism was analyzed in
detail in our recent works on noise-induced synchronization and
Liouvillian spectral coalescence
(Refs.~\cite{Bittner:2025aa,Tyagi:2024aa}), where correlated relaxation
was shown to induce spontaneous phase locking and non-Hermitian
degeneracies in two coupled dissipative oscillators.

\section{Quantum Phase Noise: From Microscopic Baths to QSDE}

Up to this point we treated $\dd\phi$ as a classical (commuting) increment and
obtained a statistics–dependent pure dephasing channel. We now upgrade
$\phi(t)$ to an \emph{operator--valued} phase generated by a quantum bath.
Let $F(t)$ be a Hermitian bath force and define the (Heisenberg) phase operator
$\phi(t)=\int_0^t F(\tau)\,\dd\tau$. The intermode coupling reads
\begin{equation}
H(t)\simeq H_0 - J K_\theta\,\phi(t)\quad\Rightarrow\quad
H_{\mathrm{int}}(t) = -J\,K_\theta\otimes F(t),
\label{eq:quantum-coupling}
\end{equation}
with $K_\theta=i(\mathcal T_\theta-\mathcal T_\theta^\dagger)$ as before.

\begin{widetext}

\subsection{Born--Markov (Gaussian) derivation.}
Assume a stationary Gaussian bath with correlation and response
\begin{equation}
C(\tau)=\tfrac12\langle\{F(\tau),F(0)\}\rangle,\qquad
\chi(\tau)=\tfrac{i}{\hbar}\Theta(\tau)\langle[F(\tau),F(0)]\rangle,
\end{equation}
and spectra $S_{FF}(\omega)=\int_{-\infty}^{\infty}C(\tau)e^{i\omega\tau}\dd\tau$,
$\tilde\chi(\omega)=\int_{0}^{\infty}\chi(\tau)e^{i\omega\tau}\dd\tau$.
To second order in $J$ (cumulant/Born expansion) and within the Markov
approximation one obtains
\begin{equation}
\dot\rho = -\,i\big[H_0 + H_{\mathrm{LS}},\,\rho\big]\;
-\;\frac{\Gamma_\theta}{2}\,[K_\theta,[K_\theta,\rho]],
\label{eq:quantum-master}
\end{equation}
with the \emph{Lamb shift}
\begin{equation}
H_{\mathrm{LS}} \;=\; J^2\,\Xi\,K_\theta^2,\qquad
\Xi \;=\; \mathrm{P}\!\int_{-\infty}^{\infty}\frac{\dd\omega}{2\pi}\,
\frac{S_{FF}(\omega)}{\omega}\;=\; \int_0^\infty\chi(\tau)\,\dd\tau,
\end{equation}
and the pure dephasing rate set by the \emph{symmetrized} zero--frequency noise,
\begin{equation}
{\quad \Gamma_\theta \;=\; 2J^2\,S_{FF}(0)\quad}
\label{eq:Gamma-quantum}
\end{equation}
(we set $\hbar=1$). Equation~\eqref{eq:quantum-master} is identical in form to
the classical result, with the replacement $D_\phi\mapsto S_{FF}(0)$.
At finite temperature $T$, $S_{FF}(0)$ carries the usual thermal factor
(e.g.\ coth$(\omega/2T)$ in an Ohmic bath), while squeezed or nonclassical baths
modify $S_{FF}(0)$ accordingly.

\subsection{QSDE (Hudson--Parthasarathy) route.}
Alternatively, couple the system to a bosonic input field and write a quantum
stochastic differential equation for the joint unitary $U_t$:
\begin{equation}
\dd U_t = \Big\{L\,\dd B_t^\dagger - L^\dagger \dd B_t
- \big(\tfrac12L^\dagger L + iH_0\big)\dd t\Big\}U_t,
\quad L=\sqrt{\gamma}\,K_\theta,
\end{equation}
with Itô table (thermal occupancy $n_{\mathrm{th}}$)
$\dd B_t\,\dd B_t^\dagger=(n_{\mathrm{th}}+1)\dd t$, $\dd B_t^\dagger\,\dd B_t=n_{\mathrm{th}}\dd t$,
others $=0$. Tracing out the field gives the Lindblad master equation
\begin{equation}
\dot\rho = -\,i[H_0,\rho]\;+\;\gamma\Big(K_\theta\rho K_\theta - \tfrac12\{K_\theta^2,\rho\}\Big),
\end{equation}
i.e.\ $-\tfrac{\Gamma_\theta}{2}[K_\theta,[K_\theta,\rho]]$ with $\Gamma_\theta=2\gamma$.
For a thermal (or squeezed) input field, $\gamma$ inherits the appropriate
quantum noise prefactors and frequency dependence. In this language the
``quantum Stratonovich'' symbol is not used; one works directly with the quantum
It{\^o} rules.

\paragraph{Multiple links and \emph{quantum} correlations.}
For links $a,b$ with operators $K_\theta^{(a)}$ and bath forces $F_a(t)$,
\begin{equation}
H_{\mathrm{int}} = -\sum_a J_a\,K_\theta^{(a)}\otimes F_a(t),\qquad
S_{ab}(\omega)=\int\! \dd\tau \, \tfrac12 \langle\{F_a(\tau),F_b(0)\}\rangle e^{i\omega\tau}.
\end{equation}
In the white--noise Markov limit,
\begin{equation}
{\;
\dot\rho = -\,i[H_0+H_{\mathrm{LS}},\rho]
-\frac12\sum_{a,b}\Gamma_{ab}\,[K_\theta^{(a)},[K_\theta^{(b)},\rho]],\quad
\Gamma_{ab}=2J_aJ_b\,S_{ab}(0),\;}
\label{eq:multi-quantum}
\end{equation}
with $H_{\mathrm{LS}}\propto \sum_{a,b}J_aJ_b\,\Xi_{ab}\{K_\theta^{(a)},K_\theta^{(b)}\}$ and
$\Xi_{ab}$ the principal--value transforms of the bath susceptibilities.
Complete positivity requires the \emph{matrix} $S_{ab}(0)$ to be positive
semidefinite (quantum Bochner theorem). If $S_{ab}(0)$ is real symmetric,
the dissipator is a sum of Hermitian--jump channels and remains normal
(no spectral EP). Complex/chiral $S_{ab}(0)$ (nonreciprocal baths, feedback,
or squeezed--phase correlations) render the dissipator non--normal and can
produce genuine decoherence exceptional points.

\subsection{Summary}
Treating $\dd\phi$ as quantum noise does not alter the \emph{structure} of the
statistics--dependent dephasing channel: we still obtain
$-\tfrac{\Gamma_\theta}{2}[K_\theta,[K_\theta,\rho]]$, but with a rate fixed by the
\emph{symmetrized quantum} zero--frequency noise $S_{FF}(0)$ and with a Lamb shift
from the antisymmetric (susceptibility) part. In multi--link settings the
quantum cross--spectrum $S_{ab}(0)$ is the fundamental object that controls rates,
correlations, and the possibility of decoherence EPs.

\end{widetext}

\section{Discussion}

Our analysis shows that the impact of stochastic exchange phases on anyonic
coherence is governed by the structure of the correlation matrix $D_{ab}$.
Because $D$ is real and symmetric, it can always be diagonalized into a set of
orthogonal eigenmodes. Each eigenvector corresponds to a collective exchange
current---or, equivalently, an irreducible representation of the system---that
couples to the environment with a rate given by the corresponding eigenvalue.
Modes associated with vanishing eigenvalues are protected from dephasing and
form decoherence-free subspaces. This establishes a clear design principle:
by engineering correlations so that $D$ develops null modes, one can guarantee
that specific collective excitations remain robust against stochastic
decoherence.

The role of $D$ is therefore to select which irreducible representations of the
anyon system couple to noise. Exceptional points in the Liouvillian spectrum
emerge when these noise-selected modes coalesce with relaxation channels or
Hamiltonian couplings, producing spectral degeneracies accompanied by
non-orthogonal eigenvectors. This mechanism is consistent with our earlier work
on noise-induced synchronization, where the eigenstructure of the correlation
matrix dictated which collective phases became locked and which decayed.

The most striking result of the present analysis is the universality of the
optimal statistical angle. In the minimal two-site model with balanced gain and
loss, we find that the protected mode always minimizes its dephasing at
$\theta^\star = \pi/2$, independent of the specific form of $D$. In other words,
the alignment between the protected mode and the eigenbasis of the
exchange-current operator $K_\theta$ occurs universally at half-fermionic
statistics. This insensitivity to noise correlations highlights a robust design
principle: regardless of how the stochastic phases are correlated, the
half-fermion point remains optimal for coherence protection.

Extensions of this framework include colored noise, such as Ornstein--Uhlenbeck
processes, which endow the rates with frequency dependence, and more general
graph geometries, where $D$ becomes the Laplacian of a correlation network. In
all cases the essential features persist: the eigenvectors of $D$ dictate which
modes are protected, while the universality of $\theta^\star = \pi/2$ provides a
simple and powerful rule for optimizing anyonic coherence.

Other recent theoretical advances have underscored the fundamental role of
statistics in determining coherence and dynamical distinguishability in
low-dimensional quantum systems. Mackel, Yang, and del Campo
\cite{2210.10776v3} demonstrated a universal orthogonality catastrophe
for one-dimensional anyons, showing that the overlap between states with
different statistical parameters decays in a manner governed solely by
the exchange statistics. Our present framework extends this notion from
a static geometric effect to a fully dynamical one: by promoting the
statistical phase $\theta$ to a stochastic variable, we derive the
resulting Lindblad dephasing channel and identify the universal
half-fermionic protection point at $\theta^\star=\pi/2$. More generally,
our treatment connects to recent studies of decoherence and spectral
universality in open quantum systems. del Campo and collaborators
\cite{PhysRevLett.119.130401} analyzed non-exponential survival
probabilities and demonstrated universal long-time scaling laws for
decoherence in complex environments. Zhao \emph{et al.}
\cite{PhysRevLett.122.014103} further showed that decoherence in generic
Markovian processes can be characterized spectrally by the eigenvalue
distribution of the Liouvillian superoperator. Within our stochastic
anyon framework, the correlation matrix $\mathsf D$ plays an analogous
spectral role: its eigenstructure selects protected and dissipative
modes, giving rise to decoherence-free subspaces and, when correlations
become complex or chiral, genuine Liouvillian exceptional points. Taken
together, these developments situate the present work within a growing
effort to unify the geometric and statistical origins of decoherence in
quantum many-body systems.

\begin{acknowledgments}
The work at the University of Houston was supported by the National Science Foundation under CHE-2404788 and the Robert A. Welch Foundation (E-1337).
\end{acknowledgments}

\section*{Data Accessibility Statement}
\textit{Computational details.} All symbolic and numerical routines (e.g., evaluation of 
$\mathrm{Var}_{|u\rangle}(K_\theta)$, search for $\theta^\star$, and lifetime estimates) 
are provided in the Supplemental Information (SI).

\section*{Author Contribution Statement}
ERB conceived the project, developed the theoretical framework,
and carried out all analytical and numerical calculations.
He wrote the manuscript, prepared the figures, and approved the final version
for submission.

\section*{Conflict of Interest Statement}
The author declares no competing interests.

\bibliography{References_local-2}

%aipnum4-2.bst 2019-01-14 (MD) hand-edited version of apsrev4-1.bst
\begin{thebibliography}{17}%
\makeatletter
\providecommand \@ifxundefined [1]{%
 \@ifx{#1\undefined}
}%
\providecommand \@ifnum [1]{%
 \ifnum #1\expandafter \@firstoftwo
 \else \expandafter \@secondoftwo
 \fi
}%
\providecommand \@ifx [1]{%
 \ifx #1\expandafter \@firstoftwo
 \else \expandafter \@secondoftwo
 \fi
}%
\providecommand \natexlab [1]{#1}%
\providecommand \enquote  [1]{``#1''}%
\providecommand \bibnamefont  [1]{#1}%
\providecommand \bibfnamefont [1]{#1}%
\providecommand \citenamefont [1]{#1}%
\providecommand \href@noop [0]{\@secondoftwo}%
\providecommand \href [0]{\begingroup \@sanitize@url \@href}%
\providecommand \@href[1]{\@@startlink{#1}\@@href}%
\providecommand \@@href[1]{\endgroup#1\@@endlink}%
\providecommand \@sanitize@url [0]{\catcode `\\12\catcode `\$12\catcode
  `\&12\catcode `\#12\catcode `\^12\catcode `\_12\catcode `\%12\relax}%
\providecommand \@@startlink[1]{}%
\providecommand \@@endlink[0]{}%
\providecommand \url  [0]{\begingroup\@sanitize@url \@url }%
\providecommand \@url [1]{\endgroup\@href {#1}{\urlprefix }}%
\providecommand \urlprefix  [0]{URL }%
\providecommand \Eprint [0]{\href }%
\providecommand \doibase [0]{https://doi.org/}%
\providecommand \selectlanguage [0]{\@gobble}%
\providecommand \bibinfo  [0]{\@secondoftwo}%
\providecommand \bibfield  [0]{\@secondoftwo}%
\providecommand \translation [1]{[#1]}%
\providecommand \BibitemOpen [0]{}%
\providecommand \bibitemStop [0]{}%
\providecommand \bibitemNoStop [0]{.\EOS\space}%
\providecommand \EOS [0]{\spacefactor3000\relax}%
\providecommand \BibitemShut  [1]{\csname bibitem#1\endcsname}%
\let\auto@bib@innerbib\@empty
%</preamble>
\bibitem [{\citenamefont {Leinaas}\ and\ \citenamefont
  {Myrheim}(1977)}]{Leinaas1977}%
  \BibitemOpen
  \bibfield  {author} {\bibinfo {author} {\bibfnamefont {J.~M.}\ \bibnamefont
  {Leinaas}}\ and\ \bibinfo {author} {\bibfnamefont {J.}~\bibnamefont
  {Myrheim}},\ }\bibfield  {title} {\enquote {\bibinfo {title} {On the theory
  of identical particles},}\ }\href {https://doi.org/10.1007/BF02727953}
  {\bibfield  {journal} {\bibinfo  {journal} {Il Nuovo Cimento B}\ }\textbf
  {\bibinfo {volume} {37}},\ \bibinfo {pages} {1--23} (\bibinfo {year}
  {1977})}\BibitemShut {NoStop}%
\bibitem [{\citenamefont {Wilczek}(1982)}]{Wilczek1982}%
  \BibitemOpen
  \bibfield  {author} {\bibinfo {author} {\bibfnamefont {F.}~\bibnamefont
  {Wilczek}},\ }\bibfield  {title} {\enquote {\bibinfo {title} {Quantum
  mechanics of fractional-spin particles},}\ }\href
  {https://doi.org/10.1103/PhysRevLett.49.957} {\bibfield  {journal} {\bibinfo
  {journal} {Phys. Rev. Lett.}\ }\textbf {\bibinfo {volume} {49}},\ \bibinfo
  {pages} {957--959} (\bibinfo {year} {1982})}\BibitemShut {NoStop}%
\bibitem [{\citenamefont {Laughlin}(1983)}]{Laughlin1983}%
  \BibitemOpen
  \bibfield  {author} {\bibinfo {author} {\bibfnamefont {R.~B.}\ \bibnamefont
  {Laughlin}},\ }\bibfield  {title} {\enquote {\bibinfo {title} {Anomalous
  quantum Hall effect: An incompressible quantum fluid with fractionally
  charged excitations},}\ }\href {https://doi.org/10.1103/PhysRevLett.50.1395}
  {\bibfield  {journal} {\bibinfo  {journal} {Phys. Rev. Lett.}\ }\textbf
  {\bibinfo {volume} {50}},\ \bibinfo {pages} {1395--1398} (\bibinfo {year}
  {1983})}\BibitemShut {NoStop}%
\bibitem [{\citenamefont {Arovas}, \citenamefont {Schrieffer},\ and\
  \citenamefont {Wilczek}(1984)}]{Arovas1984}%
  \BibitemOpen
  \bibfield  {author} {\bibinfo {author} {\bibfnamefont {D.}~\bibnamefont
  {Arovas}}, \bibinfo {author} {\bibfnamefont {J.~R.}\ \bibnamefont
  {Schrieffer}},\ and\ \bibinfo {author} {\bibfnamefont {F.}~\bibnamefont
  {Wilczek}},\ }\bibfield  {title} {\enquote {\bibinfo {title} {Fractional
  statistics and the quantum hall effect},}\ }\href
  {https://doi.org/10.1103/PhysRevLett.53.722} {\bibfield  {journal} {\bibinfo
  {journal} {Phys. Rev. Lett.}\ }\textbf {\bibinfo {volume} {53}},\ \bibinfo
  {pages} {722--723} (\bibinfo {year} {1984})}\BibitemShut {NoStop}%
\bibitem [{\citenamefont {Haldane}(1991)}]{Haldane1991}%
  \BibitemOpen
  \bibfield  {author} {\bibinfo {author} {\bibfnamefont {F.~D.~M.}\
  \bibnamefont {Haldane}},\ }\bibfield  {title} {\enquote {\bibinfo {title}
  {“Fractional statistics” in arbitrary dimensions: A generalization of the
  Pauli principle},}\ }\href {https://doi.org/10.1103/PhysRevLett.67.937}
  {\bibfield  {journal} {\bibinfo  {journal} {Phys. Rev. Lett.}\ }\textbf
  {\bibinfo {volume} {67}},\ \bibinfo {pages} {937--940} (\bibinfo {year}
  {1991})}\BibitemShut {NoStop}%
\bibitem [{\citenamefont {Kundu}(1999)}]{Kundu1999}%
  \BibitemOpen
  \bibfield  {author} {\bibinfo {author} {\bibfnamefont {A.}~\bibnamefont
  {Kundu}},\ }\bibfield  {title} {\enquote {\bibinfo {title} {Exact solution of
  double $\delta$ function Bose gas through an interacting anyon gas},}\ }\href
  {https://doi.org/10.1103/PhysRevLett.83.1275} {\bibfield  {journal} {\bibinfo
   {journal} {Phys. Rev. Lett.}\ }\textbf {\bibinfo {volume} {83}},\ \bibinfo
  {pages} {1275--1278} (\bibinfo {year} {1999})}\BibitemShut {NoStop}%
\bibitem [{\citenamefont {Kitaev}(2003)}]{Kitaev2003}%
  \BibitemOpen
  \bibfield  {author} {\bibinfo {author} {\bibfnamefont {A.~Y.}\ \bibnamefont
  {Kitaev}},\ }\bibfield  {title} {\enquote {\bibinfo {title} {Fault-tolerant
  quantum computation by anyons},}\ }\href
  {https://doi.org/10.1016/S0003-4916(02)00018-0} {\bibfield  {journal}
  {\bibinfo  {journal} {Annals of Physics}\ }\textbf {\bibinfo {volume}
  {303}},\ \bibinfo {pages} {2--30} (\bibinfo {year} {2003})}\BibitemShut
  {NoStop}%
\bibitem [{\citenamefont {Nayak}\ \emph {et~al.}(2008)\citenamefont {Nayak},
  \citenamefont {Simon}, \citenamefont {Stern}, \citenamefont {Freedman},\ and\
  \citenamefont {Das~Sarma}}]{Nayak2008}%
  \BibitemOpen
  \bibfield  {author} {\bibinfo {author} {\bibfnamefont {C.}~\bibnamefont
  {Nayak}}, \bibinfo {author} {\bibfnamefont {S.~H.}\ \bibnamefont {Simon}},
  \bibinfo {author} {\bibfnamefont {A.}~\bibnamefont {Stern}}, \bibinfo
  {author} {\bibfnamefont {M.}~\bibnamefont {Freedman}},\ and\ \bibinfo
  {author} {\bibfnamefont {S.}~\bibnamefont {Das~Sarma}},\ }\bibfield  {title}
  {\enquote {\bibinfo {title} {Non-Abelian anyons and topological quantum
  computation},}\ }\href {https://doi.org/10.1103/RevModPhys.80.1083}
  {\bibfield  {journal} {\bibinfo  {journal} {Reviews of Modern Physics}\
  }\textbf {\bibinfo {volume} {80}},\ \bibinfo {pages} {1083--1159} (\bibinfo
  {year} {2008})}\BibitemShut {NoStop}%
\bibitem [{\citenamefont {Kwan}\ \emph {et~al.}(2024)\citenamefont {Kwan},
  \citenamefont {Segura}, \citenamefont {Li}, \citenamefont {Kim},
  \citenamefont {Gorshkov}, \citenamefont {Eckardt}, \citenamefont
  {Bakkali-Hassani},\ and\ \citenamefont {Greiner}}]{Kwan2024Science}%
  \BibitemOpen
  \bibfield  {author} {\bibinfo {author} {\bibfnamefont {J.}~\bibnamefont
  {Kwan}}, \bibinfo {author} {\bibfnamefont {P.}~\bibnamefont {Segura}},
  \bibinfo {author} {\bibfnamefont {Y.}~\bibnamefont {Li}}, \bibinfo {author}
  {\bibfnamefont {S.}~\bibnamefont {Kim}}, \bibinfo {author} {\bibfnamefont
  {A.~V.}\ \bibnamefont {Gorshkov}}, \bibinfo {author} {\bibfnamefont
  {A.}~\bibnamefont {Eckardt}}, \bibinfo {author} {\bibfnamefont
  {B.}~\bibnamefont {Bakkali-Hassani}},\ and\ \bibinfo {author} {\bibfnamefont
  {M.}~\bibnamefont {Greiner}},\ }\bibfield  {title} {\enquote {\bibinfo
  {title} {Realization of one-dimensional anyons with arbitrary statistical
  phase},}\ }\href {https://doi.org/10.1126/science.adi3252} {\bibfield
  {journal} {\bibinfo  {journal} {Science}\ }\textbf {\bibinfo {volume}
  {386}},\ \bibinfo {pages} {1055--1060} (\bibinfo {year} {2024})}\BibitemShut
  {NoStop}%
\bibitem [{\citenamefont {Bittner}\ and\ \citenamefont
  {Tyagi}(2024)}]{Bittner2024}%
  \BibitemOpen
  \bibfield  {author} {\bibinfo {author} {\bibfnamefont {E.~R.}\ \bibnamefont
  {Bittner}}\ and\ \bibinfo {author} {\bibfnamefont {B.}~\bibnamefont
  {Tyagi}},\ }\bibfield  {title} {\enquote {\bibinfo {title} {Noise-induced
  synchronization in coupled quantum oscillators},}\ }\href
  {https://doi.org/10.1063/5.0246275} {\bibfield  {journal} {\bibinfo
  {journal} {The Journal of Chemical Physics}\ }\textbf {\bibinfo {volume}
  {162}},\ \bibinfo {pages} {104116} (\bibinfo {year} {2024})}\BibitemShut
  {NoStop}%
\bibitem [{\citenamefont {Bittner}\ \emph {et~al.}(2024)\citenamefont
  {Bittner}, \citenamefont {Li}, \citenamefont {Shah}, \citenamefont
  {Silva-Acuña},\ and\ \citenamefont {Piryatinski}}]{Bittner2024b}%
  \BibitemOpen
  \bibfield  {author} {\bibinfo {author} {\bibfnamefont {E.~R.}\ \bibnamefont
  {Bittner}}, \bibinfo {author} {\bibfnamefont {H.}~\bibnamefont {Li}},
  \bibinfo {author} {\bibfnamefont {S.~A.}\ \bibnamefont {Shah}}, \bibinfo
  {author} {\bibfnamefont {C.}~\bibnamefont {Silva-Acuña}},\ and\ \bibinfo
  {author} {\bibfnamefont {A.}~\bibnamefont {Piryatinski}},\ }\bibfield
  {title} {\enquote {\bibinfo {title} {Correlated noise enhancement of
  coherence and fidelity in coupled qubits},}\ }\href
  {https://doi.org/10.1080/14786435.2024.2341011} {\bibfield  {journal}
  {\bibinfo  {journal} {Philosophical Magazine}\ }\textbf {\bibinfo {volume}
  {104}},\ \bibinfo {pages} {630--646} (\bibinfo {year} {2024})}\BibitemShut
  {NoStop}%
\bibitem [{\citenamefont {Hou}\ \emph {et~al.}(2025)\citenamefont {Hou},
  \citenamefont {Ai}, \citenamefont {You},\ and\ \citenamefont
  {Zhong}}]{Hou2025prr}%
  \BibitemOpen
  \bibfield  {author} {\bibinfo {author} {\bibfnamefont {Y.-B.}\ \bibnamefont
  {Hou}}, \bibinfo {author} {\bibfnamefont {X.}~\bibnamefont {Ai}}, \bibinfo
  {author} {\bibfnamefont {R.}~\bibnamefont {You}},\ and\ \bibinfo {author}
  {\bibfnamefont {C.}~\bibnamefont {Zhong}},\ }\bibfield  {title} {\enquote
  {\bibinfo {title} {Correlated noise can be beneficial to quantum
  transducers},}\ }\href {https://doi.org/10.1103/PhysRevResearch.7.L032042}
  {\bibfield  {journal} {\bibinfo  {journal} {Physical Review Research}\
  }\textbf {\bibinfo {volume} {7}},\ \bibinfo {pages} {L032042} (\bibinfo
  {year} {2025})}\BibitemShut {NoStop}%
\bibitem [{\citenamefont {Bittner}\ and\ \citenamefont
  {Tyagi}(2025)}]{Bittner:2025aa}%
  \BibitemOpen
  \bibfield  {author} {\bibinfo {author} {\bibfnamefont {E.~R.}\ \bibnamefont
  {Bittner}}\ and\ \bibinfo {author} {\bibfnamefont {B.}~\bibnamefont
  {Tyagi}},\ }\bibfield  {title} {\enquote {\bibinfo {title} {Noise-induced
  synchronization in coupled quantum oscillators},}\ }\href
  {https://doi.org/10.1063/5.0246275} {\bibfield  {journal} {\bibinfo
  {journal} {The Journal of Chemical Physics}\ }\textbf {\bibinfo {volume}
  {162}},\ \bibinfo {pages} {104116} (\bibinfo {year} {2025})},\ \Eprint
  {https://arxiv.org/abs/https://pubs.aip.org/aip/jcp/article-pdf/doi/10.1063/5.0246275/20435238/104116\_1\_5.0246275.pdf}
  {https://pubs.aip.org/aip/jcp/article-pdf/doi/10.1063/5.0246275/20435238/104116\_1\_5.0246275.pdf}
  \BibitemShut {NoStop}%
\bibitem [{\citenamefont {Tyagi}\ \emph {et~al.}(2024)\citenamefont {Tyagi},
  \citenamefont {Li}, \citenamefont {Bittner}, \citenamefont {Piryatinski},\
  and\ \citenamefont {Silva-Acu{\~n}a}}]{Tyagi:2024aa}%
  \BibitemOpen
  \bibfield  {author} {\bibinfo {author} {\bibfnamefont {B.}~\bibnamefont
  {Tyagi}}, \bibinfo {author} {\bibfnamefont {H.}~\bibnamefont {Li}}, \bibinfo
  {author} {\bibfnamefont {E.~R.}\ \bibnamefont {Bittner}}, \bibinfo {author}
  {\bibfnamefont {A.}~\bibnamefont {Piryatinski}},\ and\ \bibinfo {author}
  {\bibfnamefont {C.}~\bibnamefont {Silva-Acu{\~n}a}},\ }\bibfield  {title}
  {\enquote {\bibinfo {title} {Noise-induced quantum synchronization and
  entanglement in a quantum analogue of Huygens' clock},}\ }\href
  {https://doi.org/10.1021/acs.jpclett.4c02313} {\bibfield  {journal} {\bibinfo
   {journal} {The Journal of Physical Chemistry Letters}\ }\textbf {\bibinfo
  {volume} {15}},\ \bibinfo {pages} {10896--10902} (\bibinfo {year}
  {2024})}\BibitemShut {NoStop}%
\bibitem [{\citenamefont {Mackel}, \citenamefont {Yang},\ and\ \citenamefont
  {del Campo}(2023)}]{2210.10776v3}%
  \BibitemOpen
  \bibfield  {author} {\bibinfo {author} {\bibfnamefont {V.}~\bibnamefont
  {Mackel}}, \bibinfo {author} {\bibfnamefont {W.}~\bibnamefont {Yang}},\ and\
  \bibinfo {author} {\bibfnamefont {A.}~\bibnamefont {del Campo}},\ }\bibfield
  {title} {\enquote {\bibinfo {title} {Quantum alchemy and universal
  orthogonality catastrophe in one-dimensional anyons},}\ }\href
  {https://doi.org/10.22331/q-2023-10-16-1044} {\bibfield  {journal} {\bibinfo
  {journal} {Quantum}\ }\textbf {\bibinfo {volume} {7}},\ \bibinfo {pages}
  {1044} (\bibinfo {year} {2023})},\ \Eprint {https://arxiv.org/abs/2210.10776}
  {arXiv:2210.10776 [quant-ph]} \BibitemShut {NoStop}%
\bibitem [{\citenamefont {del Campo}, \citenamefont {García-Calderón},\ and\
  \citenamefont {Muga}(2017)}]{PhysRevLett.119.130401}%
  \BibitemOpen
  \bibfield  {author} {\bibinfo {author} {\bibfnamefont {A.}~\bibnamefont {del
  Campo}}, \bibinfo {author} {\bibfnamefont {G.}~\bibnamefont
  {García-Calderón}},\ and\ \bibinfo {author} {\bibfnamefont {J.~G.}\
  \bibnamefont {Muga}},\ }\bibfield  {title} {\enquote {\bibinfo {title}
  {Universality of the quantum Zeno and anti-Zeno effects},}\ }\href
  {https://doi.org/10.1103/PhysRevLett.119.130401} {\bibfield  {journal}
  {\bibinfo  {journal} {Phys. Rev. Lett.}\ }\textbf {\bibinfo {volume} {119}},\
  \bibinfo {pages} {130401} (\bibinfo {year} {2017})}\BibitemShut {NoStop}%
\bibitem [{\citenamefont {Zhao}\ \emph {et~al.}(2019)\citenamefont {Zhao},
  \citenamefont {Xu}, \citenamefont {Xiang}, \citenamefont {Ramezani},
  \citenamefont {Gong},\ and\ \citenamefont {del
  Campo}}]{PhysRevLett.122.014103}%
  \BibitemOpen
  \bibfield  {author} {\bibinfo {author} {\bibfnamefont {P.}~\bibnamefont
  {Zhao}}, \bibinfo {author} {\bibfnamefont {Y.}~\bibnamefont {Xu}}, \bibinfo
  {author} {\bibfnamefont {Y.}~\bibnamefont {Xiang}}, \bibinfo {author}
  {\bibfnamefont {H.}~\bibnamefont {Ramezani}}, \bibinfo {author}
  {\bibfnamefont {J.}~\bibnamefont {Gong}},\ and\ \bibinfo {author}
  {\bibfnamefont {A.}~\bibnamefont {del Campo}},\ }\bibfield  {title} {\enquote
  {\bibinfo {title} {Extreme decoherence and quantum chaos},}\ }\href
  {https://doi.org/10.1103/PhysRevLett.122.014103} {\bibfield  {journal}
  {\bibinfo  {journal} {Phys. Rev. Lett.}\ }\textbf {\bibinfo {volume} {122}},\
  \bibinfo {pages} {014103} (\bibinfo {year} {2019})}\BibitemShut {NoStop}%
\end{thebibliography}%
\end{document}